\documentclass[
prb,
twocolumn,
superscriptaddress,
amsmath,amssymb,
aps,
]{revtex4-2}
\usepackage{appendix} 
\usepackage{graphicx}
\usepackage{dcolumn}
\usepackage{bm}
\usepackage[mathlines]{lineno}
\usepackage{braket}
\usepackage[T1]{fontenc}
\usepackage{comment}
\usepackage{xr-hyper}
\usepackage[colorlinks=true, linkcolor=black, citecolor=black, urlcolor=black]{hyperref}

\usepackage{soul,color,xcolor}

\makeatletter

\newcommand*{\addFileDependency}[1]{% argument=file name and extension
\typeout{(#1)}
\@addtofilelist{#1}
\IfFileExists{#1}{}{\typeout{No file #1.}}
}\makeatother

\newcommand*{\myexternaldocument}[1]{%
\externaldocument[supp-]{#1}%
\addFileDependency{#1.tex}%
\addFileDependency{#1.aux}%
} %end helper code

\myexternaldocument{supporting}

\begin{document}

%\title{Strain engineering of stable angles between graphene flakes on hexagonal boron nitride}
%
%\title{Lattice-moire commensurate stable angles of graphene flakes on hexagonal boron nitride}
%
%\title{Flake-edge-moir\'e aligned stable angles of heterobilayer flakes}
%\title{Lattice-moir\'e aligned stable angles of heterobilayer flakes}
%\title{Geometric twist-angle locking 
%via moir\'e-edge alignment 
%in two-dimensional materials}
%\title{Geometric locking of the moir\'e twist angle} 
%\title{Geometric lattice-moir\'e locking twist angles in heterobilayer flakes} 
\title{Geometric control of the moir\'e twist angle in heterobilayer flakes}

\author{Prathap Kumar Jharapla}
\email{prathapjharapla@g.uos.ac.kr}
\affiliation{Department of Physics, University of Seoul, Seoul 02504, Korea}
\author{Nicolas Leconte}
\email{lecontenicolas0@uos.ac.kr}
\affiliation{Department of Physics, University of Seoul, Seoul 02504, Korea}
\author{Zhiren He}
\affiliation{Department of Physics, University of North Texas, Denton, TX 76203, USA}
\author{Guru Khalsa}
\affiliation{Department of Physics, University of North Texas, Denton, TX 76203, USA}
\author{Jeil Jung}
\email{jeiljung@uos.ac.kr}
\affiliation{Department of Physics, University of Seoul, Seoul 02504, Korea}
\date{\today}

\begin{abstract}
We demonstrate a finite twist-angle stabilization mechanism in lattice-mismatched 2D heterobilayers, which results from the geometric alignment between the flake edges and its moire pattern. Using atomistic simulations of graphene on hexagonal boron nitride flakes with diameters of up to $\sim2500$~\AA, we identify robust metastable angles at $\sim 0.61^\circ$ for armchair and $\sim1.89^\circ$ for zigzag-edged flakes, tunable via in-plane heterostrain. 
This locking mechanism, which relies on energy barriers that are an order of magnitude larger than those of nearby metastable twist angles, provides a geometric route to precision twist-angle control of two-dimensional heterostructures and to understand the self-orientation of macroscopic flakes.
 
\end{abstract}

\maketitle

\section{\label{sec:level1} Introduction}

%\textit{Introduction}\textemdash
The moir\'e length, the characteristic scale of the moiré pattern, provides control over the kinetic energy of Bloch electrons in the low-energy electronic spectrum. These states can become unstable to electronic correlations when their bands are sufficiently flat~\cite{Bistritzer_2011, Andrei_2021, Mak_2022, Du_2023}. 
As a result, new physics and functional properties can be engineered, particularly in semiconducting two-dimensional materials, by controlling the relative twist between layers~\cite{Yankowitz2012-al, Dean2013-ai,Cao2018, Kim2016, Frisenda}. However, control of the twist angle is not always straightforward, and a reliable handle is required to continue the maturation of moir\'e materials in both basic science and technological applications. Twist angle control has been pursued through various techniques, including
tear-and-stack~\cite{Kim2016},
cutting-rotation-stacking~\cite{Chen_2016},
scanning probe manipulation~\cite{Ribeiro_Palau_2018,Hu_2022,Yang_2020},
mechanical bending~\cite{Kapfer_2023}, electrostatic actuation via MEMS~\cite{Tang_2024} and, theoretically, optical control using vortex beams~\cite{Zhiren2025optical}. 
%One promising additional route for precision twist-angle control may lie in exploiting the lattice mismatch between layers. 

%
Flake geometry and strain have emerged as key determinants of interlayer energetics and electronic structure in homobilayers. Finite flakes with well-defined edge terminations exhibit geometry-dependent electronic quantization and edge-localized states, and possibly shape-controlled shell and supershell patterns in the density of states \cite{Heiskanen,Heiskanen1,Bahamon}. Strain fields, whether intentionally applied or arising from relaxation, can act as effective gauge potentials that generate pseudomagnetic responses, or more generally reshape bands, including strain-induced flat features and quasi-1D channels, depending on the imposed symmetry~\cite{Neek,Bistrain,Sinner,Huder2018}. 
Angle-dependent interlayer-energy oscillations in homobilayers~\cite{Vahid2022} arise from rotation-driven changes in local stacking registry at the edges~\cite{Liu2023,Zhu_2021,PhysRevB.101.054109}. In finite flakes, the choice between zigzag and armchair edges can affect the amplitude of the barriers~\cite{Nakajima2009}. Collectively, these studies establish that geometric confinement and strain are powerful internal tuning parameters that set the registry-dependent interaction-energy landscape and continuously modulate the electronic spectrum, providing the mechanical and geometric groundwork for twist-angle stabilization and control in van der Waals homobilayers \cite{Bian,Zhou2025,Liu2023,ying2022}.
Aside from a nanoscale G/hBN study showing strain-driven twist plateaus~\cite{yangHetero}, the role of flake geometry and strain in heterobilayers, particularly for large flakes and geometry-selected alignment, remains largely unexplored.
Experimental progress in flake synthesis~\cite{c10010007, Yeh2016, Ruquan2015}, particularly the creation of size- and shape-controlled layers, expands this palette for property control to include flake geometry and edge termination, features that are critical to understanding friction, strain, and edge effects~\cite{DIEN, DIEN1, Ver, Filippov,GaoHeterostrain,Huder2018,Mespleflatband}.
In heterobilayers, both lattice mismatch and twist angle can independently tune the moir\'e length and electronic structure~\cite{PhysRevB.89.205414}, enabling flat bands and spatially varying interlayer coupling. This is exemplified by graphene on hexagonal boron nitride (hBN), a widely studied system that combines their complementary electronic properties with excellent mechanical and chemical stability. Tunable moir\'e superlattices in G/hBN have been demonstrated across a broad range of length scales, giving rise to bandgap openings, secondary Dirac points, and topological minibands~\cite{GEIM, Novoselov, Changgu, Tsetseris, Elias, Wang, Dae, Tsuyoshi, KHAN, Watanabe,Yankowitz2012-al, Dean2013-ai}.

In this manuscript, we unveil highly stable twist angles in heterobilayers composed of a flake and an extended substrate that depend on the shape of the flake and edge terminations. These small stable twist angles are driven purely by lattice mismatch and edge geometry, resulting from the tendency of the flake edge to align with the moir\'e pattern. The resulting alignment angle can be tuned by adjusting the lattice mismatch, either through suitable substrate selection or by applying in-plane heterostrain. 
We develop a geometric argument for the existence of metastable twist angles and validate it using atomistic simulations of graphene flakes on hBN. We demonstrate that heterostrain, a relative strain applied differently between the two layers, provides a practical route to precision control of the alignment angle. We offer new insights into previously unexplained experimental observation of the macroscopic self-orientation of graphene on hBN at twist angles close to $\sim 0.6^\circ$~\cite{Woods2016}, and extend these insights to other 2D heterobilayer materials.

Our manuscript is organized as follows. Section~\ref{sec:level2} describes the systems under consideration as well as the simulation methods, Sect.~\ref{sec:level3} discusses the alignment angles underpinning the angle-locking mechanism in heterobilayers and  how their values can be controlled using heterostrain. Sect.~\ref{sec:level4} briefly discusses the flake size dependence of these angles, Sect.~\ref{sec:level5} ventures into how relaxation effects may quantitatively affect our conclusions drawn from rigid systems, and Sect.~\ref{sec:level6} finally illustrates how substrate-engineering may help control the twist angle values. Sect.~\ref{sec:level7} finally summarizes our main conclusions.
While our analysis focuses mainly on heterobilayers, we present in the supporting information a similar analysis for homobilayer graphene and hBN systems. Supplemental Section I discusses the small-angle oscillations that exist both in homo and hetero-bilayers, and Section II focuses on the associated scaling laws for these oscillations. Sections III introduces a counting method to predict oscillations without energy calculations, whereas Section IV focuses on the relaxation effects in homobilayers. Further supplemental sections are cited within the manuscript.

\begin{figure}[bthp]
\centering
\includegraphics[width=\columnwidth]{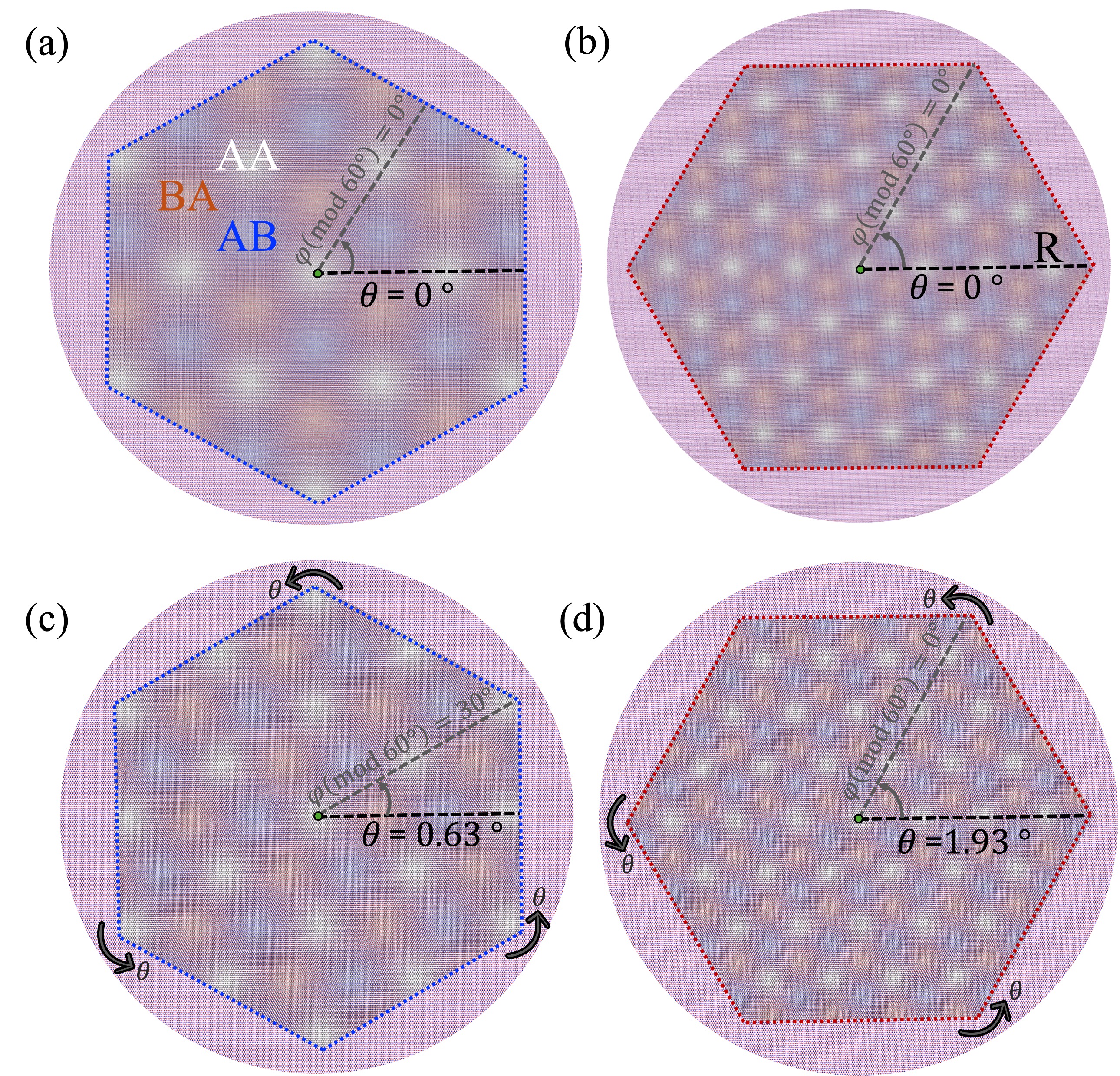}
\caption{\label{fig:fig0M}
Hexagonal graphene flakes on an extended hBN substrate, illustrating the twist angle $\theta$ and the resulting moir\'e angle $\varphi$. The dotted blue line in (a) marks an armchair edge; the dotted maroon line in (b) highlights a zigzag edge. Panels (c) and (d) show configurations at $\theta_A = 0.61^\circ$ and $\theta_Z = 1.89^\circ$, respectively, where maximal alignment between the moir\'e pattern and the flake edge is achieved, based on Eqs.~(\ref{eq:Moireangle30})~and~(\ref{eq:Moireangle60}). $R$ denotes the flake radius for both hexagonal (shown) and circular (not shown) geometries, shown here for $R= 250$~\AA.
}
\end{figure}

\section{\label{sec:level2} Systems and Methods}
%\textit{Methods}\textemdash 
To investigate this effect, we perform atomistic simulations of graphene and hBN flakes with a radius $R$ ranging from $\sim 250$~\AA\ to $\sim 2500$~\AA, using hexagonal (zigzag or armchair edges) and circular (mixed edges) geometries, as shown in Fig.~\ref{fig:fig0M}. These flakes are placed on matching substrates to form homo- (see Supplemental Material Sect.~I) or heterobilayers and are rotated by a twist angle $\theta$. Lattice constants in our calculations use pairwise interatomic potentials giving equilibrium distances of 2.46~\AA\ (graphene)~\cite{Brenner_2002} and 2.505~\AA\ (hBN)~\cite{Extep}, consistent with experimental values~\cite{donohue1982structures, SOLOZHENKO19951, 10.1063/1.1726442}. 
%Lattice constants are fixed at 2.46 Å (graphene) and 2.505 Å (hBN), consistent with the equilibrium values set by the pairwise potentials.
Interlayer distances for rigid geometries are fixed to the interlayer spacing of graphite $3.35$~\AA~  for all cases considered, and we also carry out atomic relaxation to find the equilibrium distances. 
%
%We compute both rigid, at an arbitrary interlayer distance of $3.35$\AA, and relaxed configurations. 
%
We consider both the AA- and the AB-stacking centers in the middle of the flake to position the rotation axis. We constrain one atom at the center of the graphene flake to relax only in the $z$-direction, while the other graphene atoms can relax in every direction. The hBN substrate atoms are constrained to remain rigid. The energies are computed using \textsc{LAMMPS}~\cite{LAMMPS} with EXX-RPA-inferred~\cite{Nicolas1} interlayer pair potentials~\cite{Leconte} using the DRIP functional~\cite{Wen}, and intralayer forces by REBO2~\cite{Brenner_2002} and ExTep~\cite{Extep}. The choice of intralayer potential only affects our interlayer energy observables for the relaxed systems. Energy minimizations use the conjugate gradient (CG) algorithm~\cite{POLYAK196994} with a force tolerance of $10^{-5}$ eV/\AA. The interlayer energy is computed as
\begin{equation}
E_\text{inter} = \frac{1}{2} \sum_i^N \sum_{j \notin \text{layer } i} \phi_{ij},
\label{eq:Etot}
\end{equation}
with $N=2N_F$, where $N_F$ is the number of atoms in the flake, and where pair-wise registry-dependent potential $\phi_{ij}$ is defined in Ref.~\cite{Wen}.
%In Sect.~III of the Supplemental material, we introduce a stacking-counting (SC) method of the number of AA, AB, and BA stacking centers in the flake to estimate the total energy. This SC method is simpler than existing approaches~\cite{Zhu_2019, Zhu_2021, YAN2023scalinglaw}, and allows to confirm the geometric origin of the different types of oscillations observed in our energy curves. 

\begin{figure*}
\centering
\includegraphics[width=0.8\textwidth]{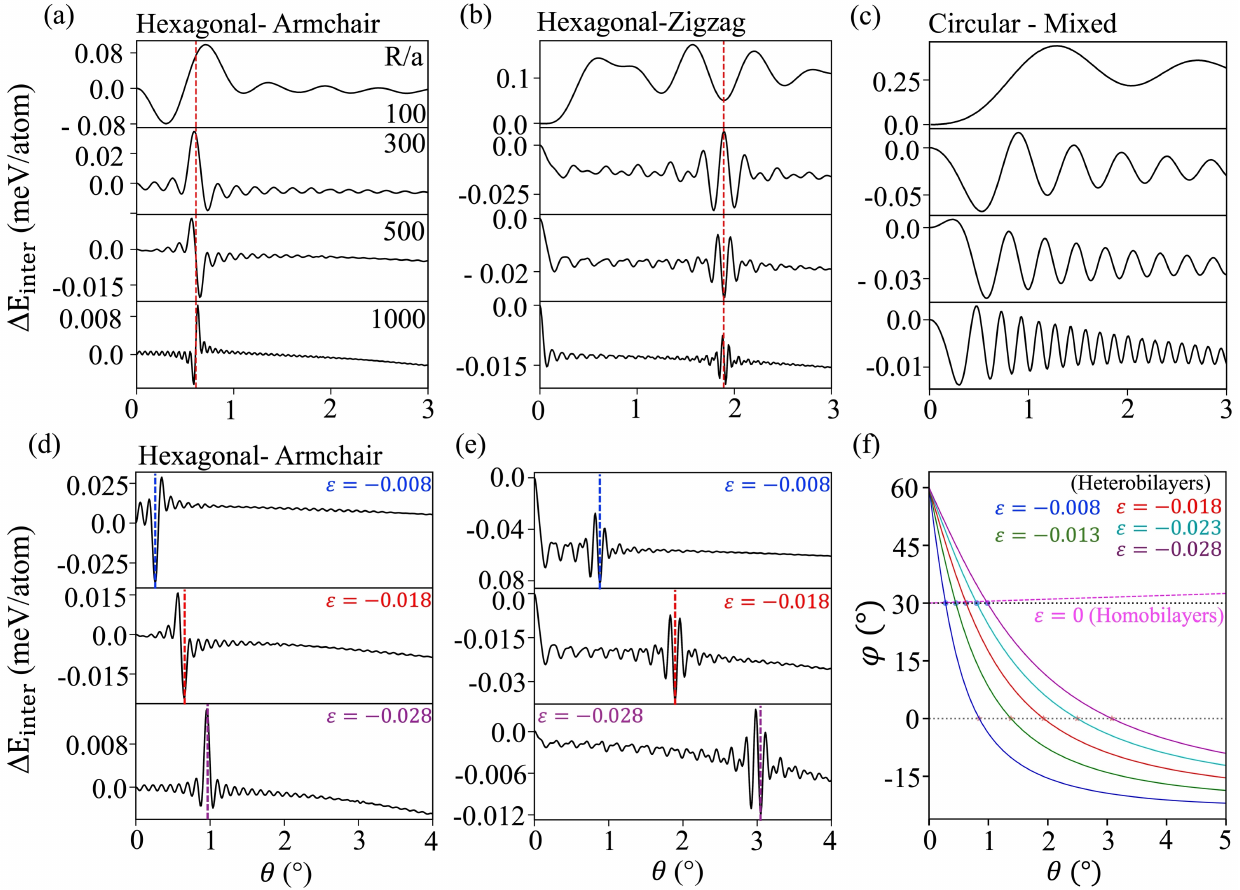}
\caption{\label{fig:fig1}
Rigid interlayer energy per atom as a function of twist angle $\theta$ for G/hBN flakes, obtained using Eq.~(\ref{eq:Etot}). Panels (a) and (b) show hexagons with armchair and zigzag edges, respectively; panel (c) shows circular flakes with mixed edges. Energies are referenced to the unrotated AA-stacked configuration. Vertical dashed lines indicate analytical predictions for alignment angles $\theta_A$ and $\theta_Z$ based on Eqs.~(\ref{eq:Moireangle30})~and~(\ref{eq:Moireangle60}). (d) and (e) illustrate how the alignment angles from (a) and (b), respectively, can be modified by changing the lattice mismatch. (f) shows the moir\'e angle $\varphi$ versus $\theta$ for different lattice mismatches $\varepsilon$. Without mismatch, $\varphi$ varies weakly at small angles, while for $\varepsilon \neq 0$, it crosses $\varphi = 30^\circ$ or $0^\circ$ at $\varepsilon$-dependent twist angles, depending on flake orientation (see Fig.~\ref{fig:fig0M}).
}
 
\end{figure*}

\section{\label{sec:level3} Alignment angles in heterobilayers}
%\textit{Lattice mismatch-induced alignment angle}\textemdash 
%
We first present the rigid interlayer energy $E_\text{inter}$ versus twist angle $\theta$ using Eq.~(\ref{eq:Etot}) for flake sizes $R/a = 100$, $300$, $500$, and $1000$, in both hexagonal and circular geometries, see Fig.~\ref{fig:fig1}. For these rigid lattice geometry calculations, only the AA-stacking rotation center configuration is considered. The lattice constant of the flake $a$ is taken here to be that of graphene. The small-amplitude oscillations in the interlayer energies in Fig.~\ref{fig:fig1} closely resemble those seen in homobilayers (see Section I and II of the Supplemental Material), with distinct maxima and minima corresponding respectively to the presence of relatively more unstable AA or more stable AB/BA stacking regions, consistent with the geometric interpretation given in Ref.~\cite{Zhu}. These angle-dependent modulations seen for rigid flakes originate from edge effects and become progressively weaker at larger twist angles due to the shrinking moir\'e length. 

Our central observation in this work is a sudden increase in the oscillation amplitudes at specific finite twist angles in armchair and zigzag edged flakes. For small flakes ($R/a < 100$), this effect is barely visible, but in larger flakes it becomes the dominant feature, overtaking the regular oscillations. Its absence in circular flakes [Fig.~\ref{fig:fig1}(c)], where only small modulations persist, suggests an origin rooted in the geometry of the flake. To confirm this, we note that these pronounced oscillations are also accurately captured by our stacking counting (SC) method, a variant on existing approaches introduced in Refs~\cite{Zhu_2019, Zhu_2021, YAN2023scalinglaw}, illustrated in Section III of the Supplemental Material,
supporting the idea that the amplified oscillations originate from the geometric alignment of the moire pattern with the flake edges. 
We illustrate in Fig.~\ref{fig:fig1}(f) the relation between the twist angle $\theta$ and the moir\'e angle~\cite{Leconte_2020}
\begin{equation}
\varphi = \tan^{-1} \left( \frac{\alpha \sin \theta}{\alpha \cos \theta - 1} \right)
\label{eq:moire_angle}
\end{equation}
where $\alpha = 1 + \varepsilon$ and $\varepsilon = (a - a_{\text{ref}})/a_{\text{ref}}$ is the lattice mismatch between $a$ and the reference lattice constant $a_\text{ref}$. In all heterobilayer cases, the substrate hBN is chosen as the reference layer. Solving Eq.~(\ref{eq:moire_angle}), we find for $\varphi=30^\circ$ 
\begin{equation}
\theta_A = \cos^{-1} \left( \frac{1 + \sqrt{-3 + 12\alpha^2}}{4\alpha} \right) = 0.61^\circ
\label{eq:Moireangle30}
\end{equation}
and, for $\varphi=0^\circ$, that
\begin{equation}
\theta_Z = \cos^{-1} \left( \frac{3 + \sqrt{-3 + 4\alpha^2}}{4\alpha}\right) = 1.89^\circ,
\label{eq:Moireangle60}
\end{equation}
corresponding to optimal alignments between the armchair~(A)- and zigzag~(Z)-terminated flakes, respectively,
and the moir\'e pattern as shown in Fig.~\ref{fig:fig0M}(c,d).
These analytical predictions, shown as dashed red lines in Fig.~\ref{fig:fig1}(f), may fall at a minimum, a maximum, or an intermediate point, depending on how the flake’s size influences the balance of stable and unstable stacking configurations along its aligned edge. Such special finite-angle alignment does not happen in homo-bilayers, because $\varphi=30^\circ$ alignment is only achieved at $\theta=0^\circ$ when using the orientation of the flake from Fig.~\ref{fig:fig0M}(a), as illustrated by the pink dashed line in Fig.~\ref{fig:fig1}(f). Indeed, we see that the moir\'e angle varies too slowly with the twist angle to produce finite-angle alignment effects in the small-angle regime (magenta line), where the coupling between the two layers is sufficiently large to have a meaningful impact on the energetics of the system.
\begin{figure}[tbph]
\centering
\includegraphics[width=1\columnwidth]{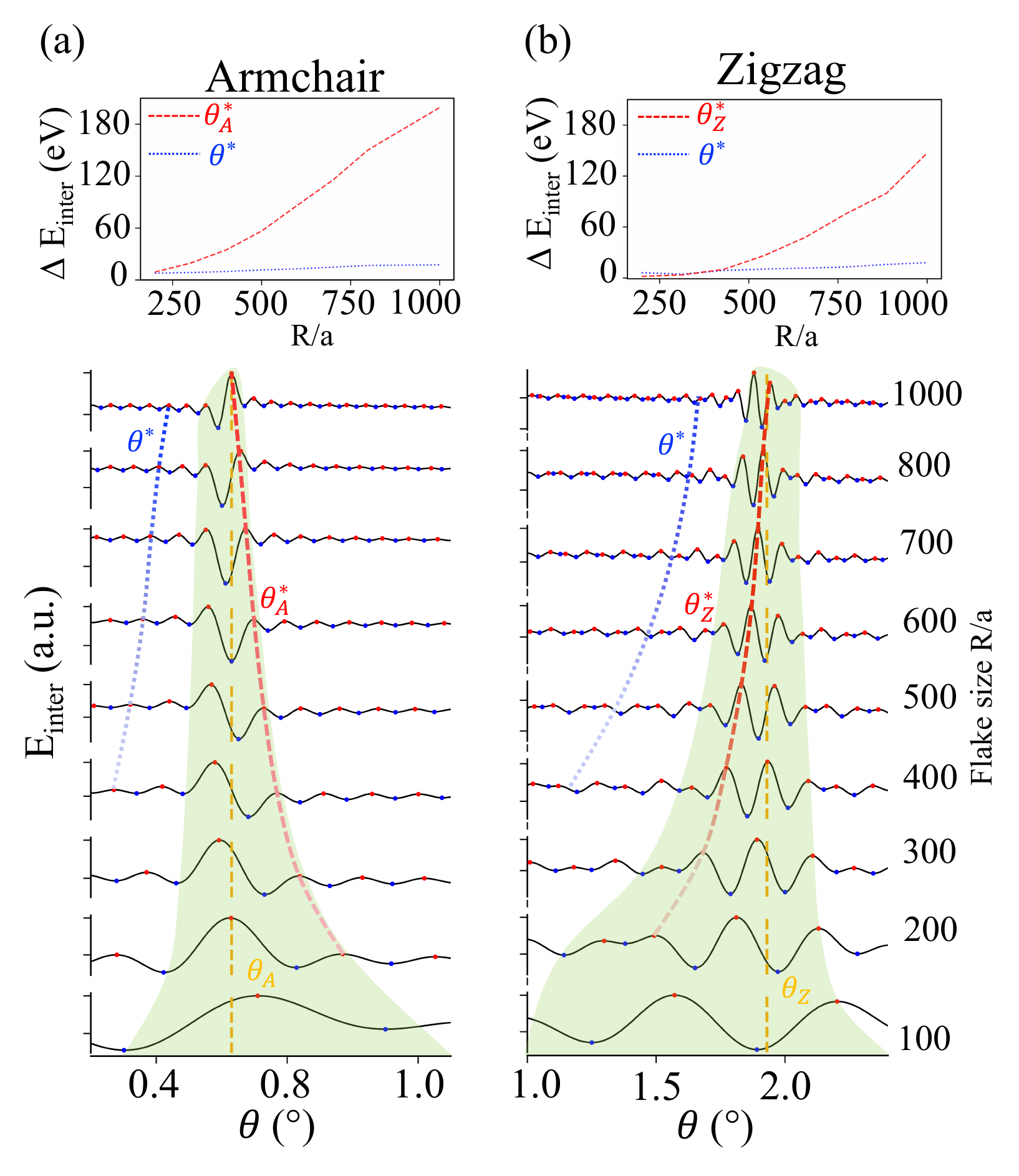}
\caption{\label{fig:fig3}
Twist-angle-dependent energy oscillations near the alignment angles $\theta_A$ and $\theta_Z$ for various flake sizes $R/a$, for armchair- (a) and zigzag-terminated (b) hexagonal flakes. The rotation center axis is chosen to lie at the AA stacking. Bottom panels show how these angles converge toward the analytical predictions (vertical dashed orange lines) as flake size increases, confirming the robustness of geometric locking. The green-shaded region emphasizes this trend. Arbitrary units are used for visual clarity. The top panels present the scaling behavior of the energy barrier, $\Delta E_\text{inter}(\theta) = E_\text{inter}(\theta) - E_\text{inter}(\theta^-)$, 
evaluated at $\theta^\star_A$ and $\theta^\star_Z$ whose maxima are close to the analytical values for $R/a=1000$ flakes.
These angles are highlighted by red curved dashed lines in the bottom panels, where $\theta^-$ denotes the adjacent energy minimum. The comparatively much smaller barrier scaling at a nearby angle extremum away from alignment (blue), labeled $\theta^*$, highlights the enhanced stability of the alignment angles.
}
\end{figure}

Since the moire alignment angles for armchair and zigzag edged flakes $\theta_A$ and $\theta_Z$ depend on the lattice mismatch, we demonstrate in Fig.~\ref{fig:fig1}(d,e) how in-plane heterostrain allows precise tuning of these angles. This is achieved by varying the flake's lattice constant, which, in the rigid regime, is equivalent to homogeneously straining the substrate. As expected, compressive strain shifts the alignment angles to smaller values, while tensile strain makes them larger. The analytical predictions in Fig.~\ref{fig:fig1}(f) closely match the numerical results shown as dashed vertical lines in panels (d,e).

\section{\label{sec:level4} Flake size-dependence of alignment angle}
%\textit{Flake size-dependence of alignment angle}---
Because the stability of the alignment angle depends on flake size and strain, the bottom panels of Fig.~\ref{fig:fig3}(a,b) show the evolution of the total energy as a function of twist angle and flake size, revealing a progressive convergence of the oscillation peaks toward the analytical alignment angles in Eqs.~(\ref{eq:Moireangle30}) and~(\ref{eq:Moireangle60}) as the flake size increases.
To quantify this trend, we define $\theta^\star_A$ and $\theta^\star_Z$ local maxima angles that lie closest to the analytical values of $\theta_A$ and $\theta_Z$ for $R/a = 1000$. Their evolution with flake size is traced by the red dashed lines in Fig.~\ref{fig:fig3}, while the green-shaded region highlights a broader range of equivalent angles, illustrating how the angles yielding large-amplitude oscillations progressively collapse onto the analytical predictions.
While the details of the energy oscillation landscape is expected to vary depending on flake size or choice of rotation center, we expect a similar convergence of amplitude maxima towards the analytical values in the large flake limit. 
This demonstrates the robustness and predictive power of the geometric model in describing twist-angle locking.
The top panels show the scaling with flake size of the energy barrier corresponding to the most prominent barriers near $\theta_A$ and $\theta_Z$ that are identified in our simulation sample, illustrating that the barrier for armchair-terminated flakes is approximately 30\% higher than for zigzag-terminated flakes, reflecting their stronger geometric locking tendency. Importantly, the energy barriers near $\theta_A$ and $\theta_Z$ (red) exceed those of a nearby smaller angle extremum, $\theta^*$ (blue), by an order of magnitude, highlighting the exceptional stability of the alignment angles. We note that energy barriers in heterobilayers are capped by the finite moir\'e length (e.g., $\sim13.5$\,nm for aligned G/hBN), unlike in homobilayers, where, theoretically, they can diverge as $\theta \to 0$. For example, at $\theta = 0.1^\circ$, $L_m$ in twisted bilayer graphene (tBG) exceeds that of G/hBN by more than an order of magnitude, leading to much larger barrier amplitudes. Near $0.6^\circ$, the tBG barriers are still about twice as large as the most prominent ones of G/hBN. For a more detailed analysis of the scaling of rotational alignment barriers with $R$, we refer to Section II of the Supplemental Material, where it is shown to deviate from the $R^2$ scaling of small-angle oscillations energy barriers.

\begin{figure}[tbhp]
\centering
\includegraphics[width=1.0\columnwidth]{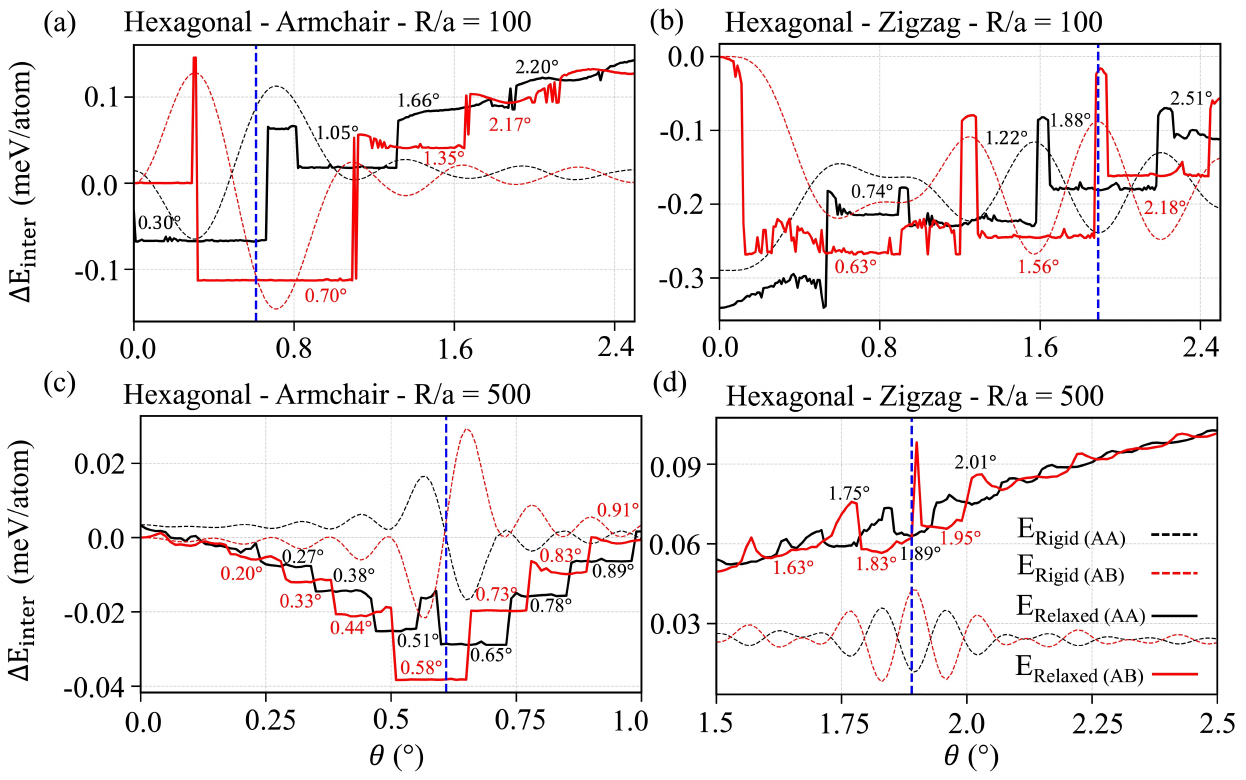}
\caption{\label{fig:fig5}
Interlayer energy $\Delta E_\text{inter} (\theta) = E_\text{inter}(\theta) - E_\text{inter}^\text{AB}(0)$ as a function of twist angle for flakes with $R/a = 100$ and $500$ for AA-stacking rotation center (black) and AB-stacking center (red). Dashed lines correspond to rigid configurations; solid lines show relaxed energies. Each curve is individually referenced to its reference $E_\text{inter}^\text{AB}(0)$ for rigid or relaxed geometries at zero twist angle, see Table~II in Sect.~VI of the Supplemental material. 
Plateaus in the relaxed curves indicate spontaneous rotation into nearby metastable angles. Final angles, extracted using the Kabsch algorithm~\cite{kabsch1976solution}, are labeled in red above each plateau. The dashed blue vertical lines indicate the analytical predictions of $\theta_A$ and $\theta_Z$. AB curves are overall more stable than corresponding AA curves.
}
\end{figure}

\section{\label{sec:level5} Lattice relaxation effects}
While the total energies for rigid flake geometries are useful to capture the qualitative behavior or total energies, allowing the atoms to relax provides a more realistic description of the twist angle locking near the energy minimum. For deeper insight on the atomic relaxation effects we set an initial twist angle and allow the total energy to relax. The total energy results are shown in Fig.~\ref{fig:fig5} compares small flakes (top panels), which do not clearly resolve alignment angles, with intermediate flakes (bottom panels) where rigid energy minima are well developed. We examine both the AA-stacking–centered configurations (black) calculated for rigid geometries in the previous section, and also AB-stacking–centered ones (red), allowing us to assess whether a flake can slide into a more energetically favorable state.
For small flakes, rigid energy oscillations (dashed) flatten into plateaus upon relaxation (solid) for both armchair (a) and zigzag (b) cases. These plateaus indicate spontaneous rotation into the same final angle, which corresponds to the angle identified by the nearest local minimum in the rigid curve. These angles, labeled on top of each constant-energy plateau, are extracted via the Kabsch algorithm~\cite{kabsch1976solution}, which minimizes the root mean square deviation between the pre- and post-relaxation atomic configurations. Equivalent relaxation results for homo-bilayers are presented in Sect.~IV of the Supplemental Material. Residual fluctuations stem from local rotational variations, as documented in section V of the Supplemental Material.
For flakes with $R/a = 500$, relaxation yields well-defined plateaus in the armchair case, where the alignment angle becomes a global minimum, indicating a strong tendency toward twist locking. In the zigzag case [panel (d)], similar plateaus emerge, but the alignment angle near $1.9^\circ$ does not coincide with a global minimum due to the increase in background energy at larger angles~\cite{PhysRevB.110.024109}. We omit large flakes (e.g., $R/a_\text{ref}=1000$) from our analysis, as their relaxation is computationally costly and, in the homobilayer case, they show no clear alignment plateaus, likely due to insufficient reorientation torque. AB-centered configurations, colored in red, are more favored in our example cases, especially at the alignment angle, supporting the assumption that the system will likely slide into this configuration at finite temperature. Their comparison also illustrates that the analytical prediction, which ignores whether perfect alignment corresponds to a local maximum or minimum in the rigid curves, is an increasingly accurate approximation of the final Kabsch angle matching perfect alignment for larger flake sizes.
%At these larger sizes and angles, the torque is insufficient, within zero-temperature simulation assumptions, to overcome barriers and induce metastable reorientation.

\begin{figure}[tbhp]
\centering
\includegraphics[width=0.8\columnwidth]{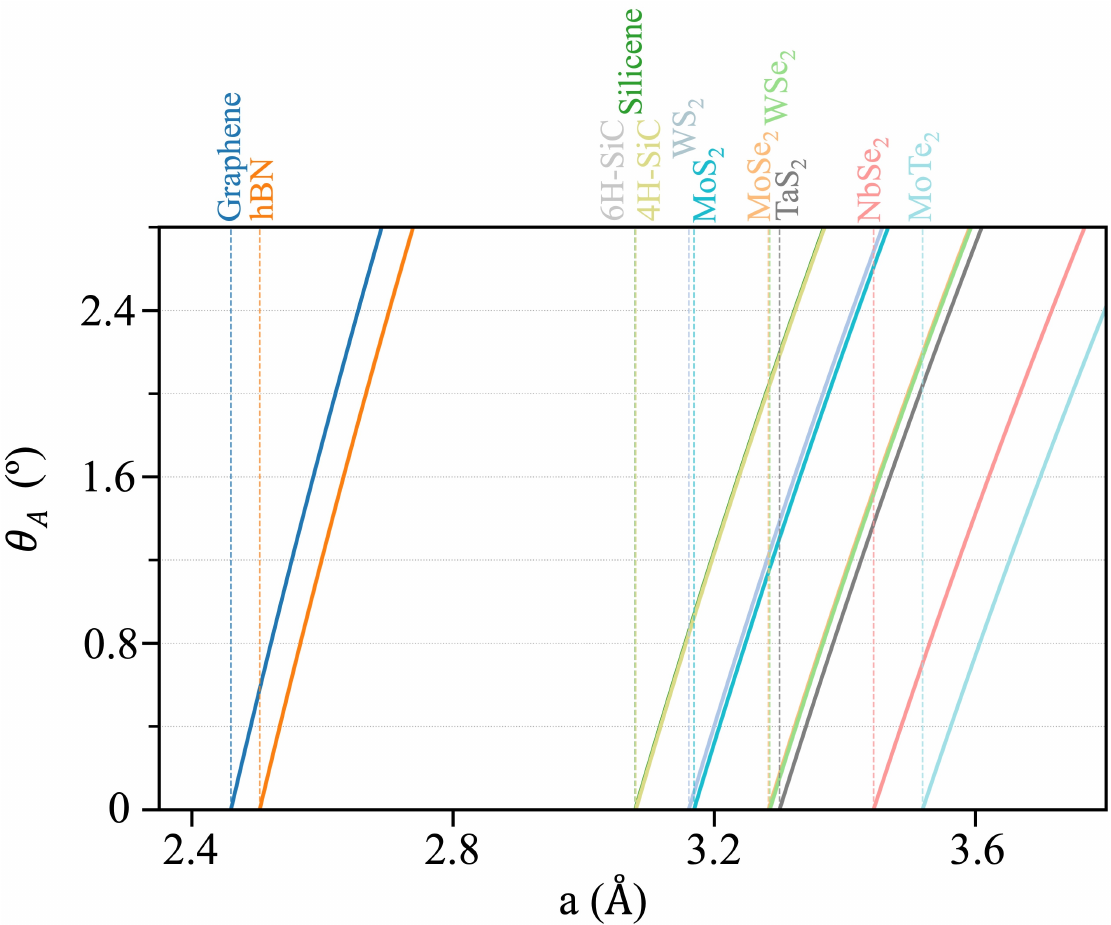}
\caption{\label{fig:fig6}
Alignment twist angle $\theta_A$ as a function of lattice constant $a$ for various 2D materials stacked on a set substrate, valid for armchair edged flakes based on Eq.~(\ref{eq:Moireangle30}). Curves indicate the angles at which each material forms an aligned moir\'e superlattice. Vertical dashed lines and color-coded labels mark experimental lattice constants of various materials, from graphene ($a = 2.456$\AA) to MoTe$_2$ ($a = 3.521$\AA), taken from the ICSD~\cite{ICSD}. Colors distinguish materials for ease of comparison. A wider palette of materials is illustrated in the Supplemental Material in Sect.~VII.
}
\end{figure} 

\section{\label{sec:level6} Substrate-driven angle engineering}
%\textit{Substrate-driven angle engineering}
Building on our G/hBN analysis of flake–substrate alignment, Eq.~(\ref{eq:Moireangle30}) provides a predictive framework for selecting material pairs that exhibit stable alignment at small twist angles, a regime of particular interest for realizing correlated electronic phases. We illustrate this in Fig.~\ref{fig:fig6} for a selection of common 2D materials. Curved lines show the alignment angle $\theta_A$ as a function of the lattice constant of the substrate. Vertical dashed lines indicate the experimental lattice constants of various materials from the ICSD~\cite{ICSD}. The intersection of a curved line with a vertical dashed line marks the expected twist angle at which the flake geometry aligns with the moir\'e pattern. This construction highlights that in flake–substrate systems, selecting materials with nearly matching lattice constants can lead to robust, self-aligned twist angles, without active twist control. Our current analysis thus provides a systematic geometric route to achieve twist angle control in pre-assembled heterobilayers
and justifies the possibility of rotation control through thermal annealing~\cite{doi:10.1126/science.aad2102} or by optical vortex beams~\cite{Zhiren2025optical}.
In contrast to conventional rotational epitaxy where twist angles are achieved during the material growth process~\cite{doi:10.1021/acs.nanolett.9b01565, PhysRevLett.41.955} and where interfacial energy minimization during growth selects discrete orientations, our mechanism originates from the commensurability between flake edges and the moiré lattice, offering deterministic and strain-tunable twist locking even in static flakes.

%\section{\label{sec:level7} Conclusions}
%\textit{Conclusions}\textemdash 
%We have demonstrated that stable twist angles in graphene flakes on hBN arise largely from the geometric alignment between flake edges and the moir\'e lattice, a mechanism not present in homobilayers. Atomistic simulations reveal sharp stability angles near $0.61^\circ$ (armchair) and $1.89^\circ$ (zigzag), corresponding to optimal edge-moir\'e alignment and enhanced rotational barriers. These angles are tunable via in-plane heterostrain, providing a practical route for twist-angle control in van der Waals heterostructures.
%
%Our geometric framework complements recent advances in flake synthesis~\cite{c10010007, Yeh2016, Ruquan2015} and explains long-standing experimental observations of macroscopic self-alignment in heteroflakes~\cite{doi:10.1126/science.aad2102, Woods2016}. While a small-angle minimum near $0.6^\circ$ has been reported in G/hBN~\cite{Woods2016}, its origin remained unclear. We can now attribute this behavior to a bulk lattice matching effect between the moir\'e lattice 
%{\bf and the atomic lattice that mirrors the edge alignment mechanism we identify in finite flakes. }
%Building on existing progress in flake synthesis, our work underscores the potential of flake geometry and heterostrain as powerful tools for engineered twist locking, encouraging further innovation in the moir\'e materials community toward realizing strongly correlated regimes without relying on active twist manipulation.

\section{\label{sec:level7} Conclusions}
We have shown that stable twist angles can arise in lattice-mismatched two-dimensional heterobilayers from a purely geometric alignment between a finite flake’s edges and its moiré pattern. Using large-scale atomistic simulations of graphene flakes on hexagonal boron nitride, supported by analytical modeling, we identified robust metastable twist angles near $0.61^{\circ}$ (armchair) and $1.89^{\circ}$ (zigzag) that result from lattice-mismatch–driven edge–moire commensurability. Our geometric framework complements recent advances in flake synthesis~\cite{c10010007, Yeh2016, Ruquan2015} and explains long-standing experimental observations of macroscopic self-alignment in heteroflakes~\cite{doi:10.1126/science.aad2102, Woods2016}. 
We can attribute this behavior to a bulk lattice matching effect between the moir\'e lattice 
and the atomic lattice, in addition to the edge alignment mechanism in finite flakes.
We provide a natural explanation for previously observed self-alignment of graphene on hBN near $0.6^{\circ}$~\cite{Woods2016} and reveal a deterministic route to twist-angle control that does not rely on external manipulation.

Our framework links the locking angle directly to lattice mismatch and edge orientation, predicting that heterostrain can tune these angles continuously. Our calculations confirm that moderate in-plane strain shifts the locking angles, offering practical means to engineer twist alignment. The locking becomes stronger with increasing flake size, with energy barriers at the alignment angles an order of magnitude larger than nearby extrema, particularly for armchair-terminated flakes, demonstrating the robustness of this effect. Including atomic relaxation preserves these stable orientations, which appear as extended plateaus in the relaxed energy curves.

The same geometric principle applies broadly to other van der Waals heterostructures: by selecting materials with specific lattice constants or applying controlled strain, one can realize reproducible, finite-angle alignment analogous to rotational epitaxy but in pre-assembled flakes. This geometry-assisted twist locking provides a static and strain-tunable approach to precision moiré engineering, complementing existing mechanical or optical control methods. Our results thus establish flake geometry and lattice mismatch as powerful system parameters for achieving reliable twist control in 2D materials heterobilayers and for designing next-generation moiré systems with controllable orientation.

%\textit{Acknowledgments}\textemdash 
\begin{acknowledgments}
This work was supported by the Korean NRF through Grant NRF-2020R1A5A1016518 (P.K.J. and J.J.) and Grant RS-2023-00249414 (N.L.).
We acknowledge computational support from KISTI Grant No. KSC-2022-CRE-0514 and by the resources of Urban Big data and AI Institute (UBAI) at UOS.
\end{acknowledgments}

\bibliographystyle{apsrev4-2}
\bibliography{aps}

%\pagebreak
%\newpage
\appendix
\setcounter{figure}{0}
\renewcommand{\thefigure}{A\arabic{figure}}

\end{document}